# Millisecond insect tracking system

T. Thang Vo-Doan and Andrew D. Straw

*Abstract*— Animals such as insects have provided a rich source of inspiration for designing robots. For example, animals navigate to goals via efficient coordination of individual motor actions, and demonstrate natural solutions to problems also faced by engineers. Recording individual body part positions during large scale movement would therefore be useful. Such multi-scale observations, however, are challenging. With video, for example, there is typically a trade-off between the volume over which an animal can be recorded and spatial resolution within the volume. Even with high pixel-count cameras, motion blur can be a challenge when using available light. Here we present a new approach for tracking animals, such as insects, with an optical system that bypasses this tradeoff by actively pointing a telephoto video camera at the animal. This system is based around high-speed pan-tilt mirrors which steer an optical path shared by a quadrant photodiode and a high-resolution, high-speed telephoto video recording system. The mirror is directed to "lock on" to the image of a 25-milligram retroreflector worn by the animal. This system allows high-magnification videography with reduced motion blur over a large tracking volume. With our prototype, we obtained millisecond order closed-loop latency and recorded videos of flying insects in a tracking volume extending to an axial distance of 3 meters and horizontally and vertically by 40 degrees. The system offers increased capabilities compared to other video recording solutions and may be useful for the study of animal behavior and the design of bio-inspired robots.

## I. INTRODUCTION

Insect locomotion has been a model for bio-inspired robotics for decades [1]. Micro- to pico- aerial vehicles (MAV, PAV) often seek to mimic mechanisms of insect flight [2]–[6], ground robots find inspiration from insect jumping, walking and running [7]–[10]. However, the study insect maneuvering is typically limited to high resolution recording of restrained insects [11]–[13] or to a small volume of interest [14]–[17]. Although existing tracking systems can track over large volumes, they have difficulty resolving body parts such as antennae, legs, and wings [2], [11]. A tracking system with the capability of imaging an insect at high resolution while following it over large distances would be a useful innovation. Oku et al. [18], [19] demonstrated the "saccade mirror" system, which keeps the gaze of a telephoto camera on a moving object within large field of view. The system relies on galvanometer motors to steer mirrors to follow the object position, which is calculated by a FPGA image processor based on the image observed by a camera viewing images through the mirror. However, using specialized low-latency

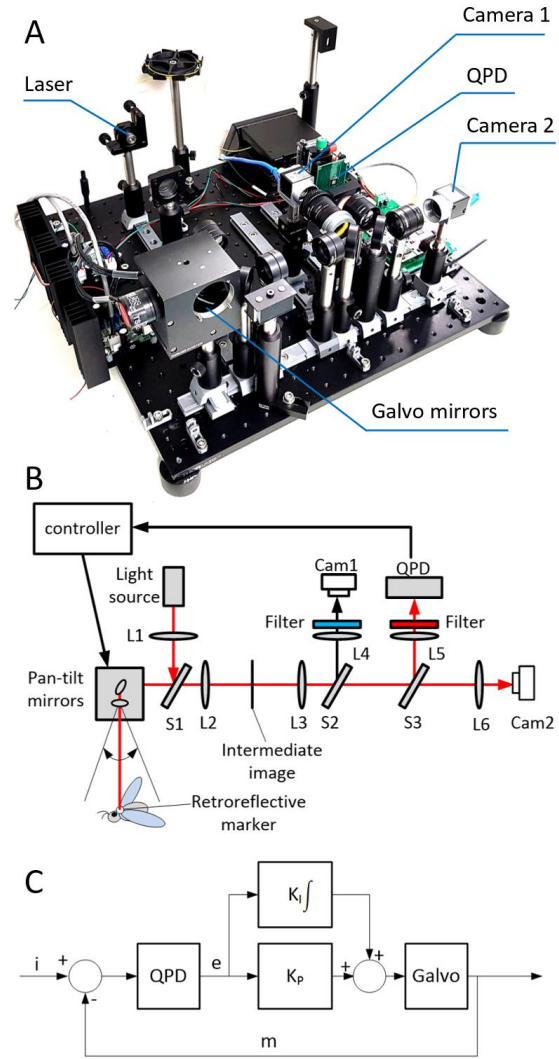

Figure 1. Overview of insect tracking system. A. The prototype assembly. B. The system schematic. The system includes pan-tilt galvo mirrors, a laser source, a QPD, camera 1 for high resolution imaging of the insect, camera 2 for debugging beam position, lenses and spliters. A high-speed controller processes analog data from the QPD and commands the galvo motors. A retroreflective marker mounted on the insect creates a spot whose image falls on the QPD. C. Block diagram of the control system. Difference between insect position (i) and mirror postiion (m) results in error (e) output by the QPD and used as input to a PI controller to generate new mirror angle, resulting in the image of the insect being centered on the QPD.

image sensors and processing is costly, both financially and technically.

In this paper, we propose a low latency tracking system that uses the analog output of a quadrant photodiode (QPD) as an angular position sensor used to control galvanometer motors. While using different actuators, similar concepts have been employed to track bacteria [20] and related ideas are employed in optical tweezers [21] and analog signal recovery in CD players [22]. Here we seek to use this mechanism to track insects at the macroscopic scale. By eliminating digital image acquisition and processing in favor of analog processing, we aimed to meet or exceed the latency performance of the digital image-based systems with reduced cost and complexity.

In this report, we document a system that tracks a moving insect at millisecond-order latency within a 40° field-of-view in both horizontal and vertical directions. High spatio-temporal resolution video was recorded during tracking using a high-speed digital telephoto camera sharing the optical path of the QPD sensor. With this system, we recorded a jump-land-jump sequence of a locust and the take-off and flight of a beetle. This tracking system enables recording video of animal behavior at high resolution over large volumes and could also be applied to track other moving objects such as robots.

## II. OPTICAL CONFIGURATION

Here we describe the optical design of our system (Figure 1a, b). The optical components consist of two mirrors arranged in pan-tilt configuration and actuated by galvanometer motors (Thorlabs, GVS212), two through the mirror (TTM) cameras (Basler acA1300-200um and acA720-520um), a quadrant photodiode (QPD, TT Electronics OPR5911), a red laser light source (637 nm, 70 mW, Thorlabs, LP637-SF70), lenses, beam splitters, optical filters and a retroreflective marker. The light source is used for illuminating the marker, which is mounted on an insect. The laser light is expanded via lens L1 and directed to the galvo mirrors by 50/50 splitter S1. The mirrors steer the optical path, including both outgoing and reflected incoming light, to follow the marker via closed loop feedback. An intermediate image of the marker and insect is formed by L2 at a position between L2 and L3. This image is refocused on the TTM cameras and QPD via lens pairs L3-L4, L3-L5, and L3-L6. As the spot-like image of the marker moves on the QPD, the photocurrent from each segment changes. The QPD output is used to steer the galvo mirrors to keep the marker centered on the QPD (details below). The optical axes of QPD and cameras are aligned. A notch filter (blocking 637nm light) is placed in front of camera 1, the insect view camera, to block the strong reflected light spot. This camera thus records images of the insect using ambient illumination. A bandpass filter (600-650 nm) is placed in front of the QPD to minimize the effects of ambient light. Camera 2 has no filter and records an image of the marker similar to that falling on the QPD for debugging.

## III. QPD CIRCUIT, FEEDBACK CONTROLLER AND CALIBRATION

Here we describe the QPD and feedback controller circuitry (Figure 1c). From control perspective, the system can be conceived as follows. The QPD sensor measures the difference of insect position i from pan-tilt mirror position m. This error angle e (equal to i-m) is input to a PI controller commanding the mirrors. In more detail, photocurrents from the QPD are fed through transimpedance amplifiers, summed and differentiated to result in two analog voltages $v_{LR}$ and $v_{TB}$ signaling the error between the spot center relative to the QPD left-right and top-bottom axes (custom circuit board design in supplemental data). These voltage signals are sampled with the onboard ADCs of a microcontroller (STMicroelectronics, STM32F103RB). On the microcontroller, a two-dimensional error e is computed. The error dimensions are aligned with the axes of the pan-tilt mirrors using the results of a calibration procedure (see below). A proportional-integral (PI) controller is implemented in the microcontroller for each mirror and generates a command voltage (DAC714, Texas Instruments) to drive each galvo axis (source code in supplemental data).

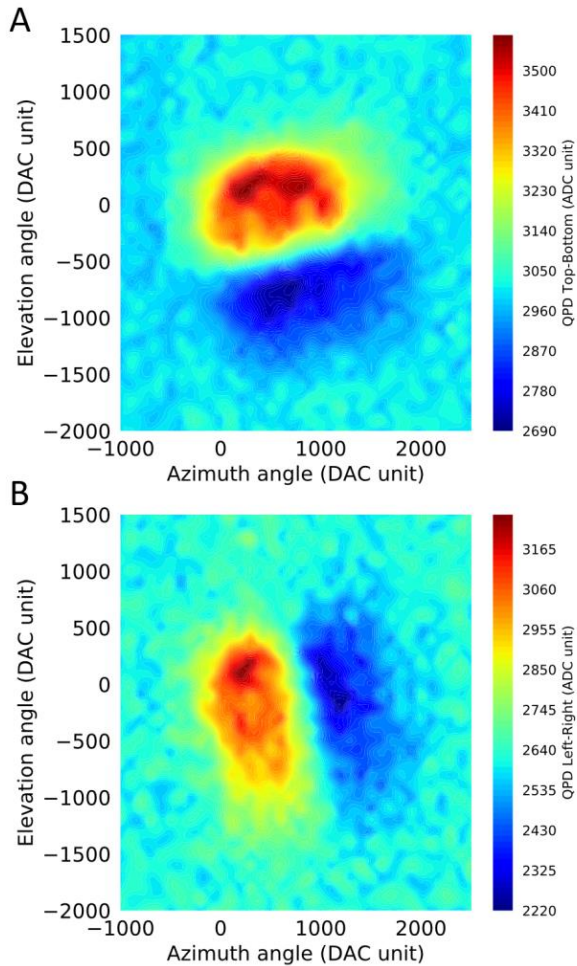

Figure 2. QPD output recorded during an open-loop scan of a stationary marker. (A) top-bottom and (B) left-right differential values as recorded by the contoller. Such calibration data is used to generate a QPD-to-angle function. When operated in closed loop, the system maintains the marker image spot in the central, near-linear region.

The voltage-to-angle transformation uses a calibration performed as follows. The optical axis of the system is directed to a given pan and tilt angles with DAC1 and DAC2 commands, respectively. Voltages $v_{LR}$ and $v_{TB}$ are recorded as ADC1 and ADC2, respectively (Figure 2). Two linear functions are generated based on the data from the central region of the scan where the results are approximately linear and which compute error angle in the pan-tilt axes in DAC units. These error angles are used as the inputs to the PI controller (calibration source code in supplemental data).

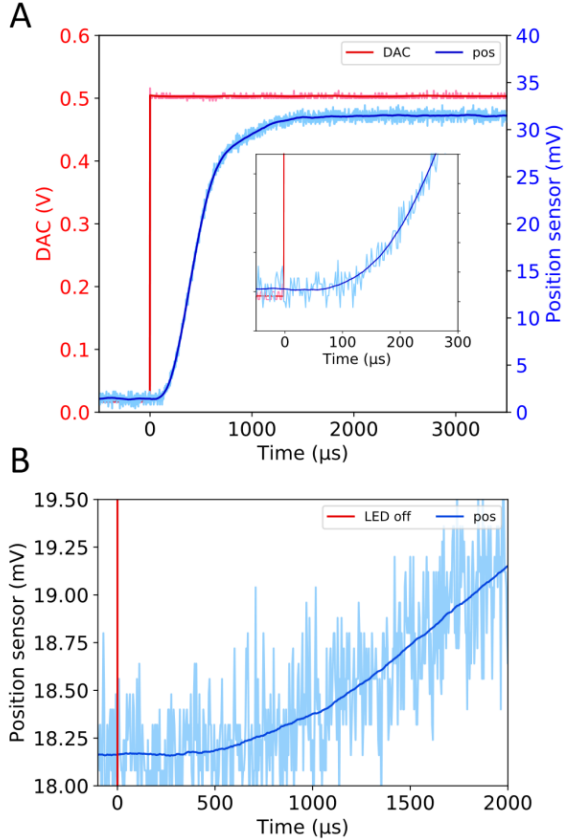

Figure 3. Latency measurements. A. Galvo and mirror latency. The response of the galvo motor and mirror during a 1 degree open-loop step. The inset shows the movement initiation period. Red line is the galvo command signal. B. Total system latency from photon to motion initiation. Red line is the LED off time. In both panels, the light blue line shows the raw position sensor output and the dark blue line shows the result of polynomial smoothing.

## IV. LATENCY EVALUATION

First, we recorded the step response of the galvanometer mirror to ensure it is applicable for low latency tracking. Figure 3 shows that the motor has a latency to initial movement of about 100 microseconds and settling time of about 1.5 milliseconds for a 1° step and is suitable for our needs. Next, we evaluated the photon-to-motion latency of the complete system including QPD sensor, controller, galvanometer motors. We did this by switching off an LED which was being tracked and recording the mirror position as the system drifted away. This measurement is therefore an upper bound of the latency of the system. Figure 3 shows that this upper bound of system latency is about 500 microseconds.

## V. INSECT TRACKING EXPERIMENT

Recording high-resolution videos of insect maneuvering is challenging due to the simultaneous requirements of sufficient resolution to see individual body parts but also a volume large enough to cover the entire region of interest. We used our prototype tracker to record high-speed video of a locust during jumping and a beetle in flight.

### A. Experiment setup

We used locusts (*Schistocerca gregaria* Forsskål 1775, Zoo Burkart, Germany) and beetles (*Pachnoda marginata*, gift from a local breeder). Retroreflective tape (3M, 8850) was used to cover the thorax and the anterior of the head of the locust or the pronotum and thorax of the beetle to make them visible to the tracking system. In addition, a 3mm retroreflective maker was mounted to the thorax of the insect to increase its visibility especially from the back. Total marker weight was 25 milligrams. Theater stage lights were used to provide ambient illumination for high-speed videography. The system is able to track the insect within the galvanometer scanning angle of 40° at a distance of 3 meters. Videos were recorded at 169 frames per second. An additional camera (GoPro, Hero3) was used to record an overview of the experiment. The locust jumping and beetle take-off were triggered manually by the experimenter. The system was set to lock onto the insect before each recording. The trajectory of the insects was extracted using DLTdv8a[23].

### B. Experimental Results

The videos recorded with the tracking system show detail including antennae and leg position at high frame rate throughout large scale maneuvering (Figures 4, 5 and Supplemental Video). Compared to the overview camera, we note that motion blur on the telephoto tracking camera is strongly reduced because of the near constant small angular offset from the maker and insect.

Three consecutive jumps of the locust were tracked and recorded on the TTM and overview cameras (Figure 4 and Supplemental Video). The system maintained a tracking "lock" even when the locust reached peak acceleration. Based on trajectory extraction from the overview camera, we estimate the accelerations to be over 20 m/s$^2$ at speeds of 1.9 m/s. While both the overview and TTM cameras capture the body of the locust, other parts like limbs and wing are blurry in the overview camera. At the first jump, the locust performed a full rotation around the pitch axis before landing and recovery. The tracking system maintained a lock on the locust when it fell at 3.5 m/s.

The system also performed stable tracking for a take off and flight sequence of a beetle and lost tracking when the beetle passed out of the tracking mirror angular limits (Figure 5 and Supplemental Video). The beetle performed a non-jumping take off. It raised its body off the ground and started to flap the wings without throwing the body into air. After taking off, it accelerated to around 2 m/s when facing unstable flight as the left wing hit the ground and could not open fully. It then flew out of the tracking volume at a peak speed of about 4 m/s.

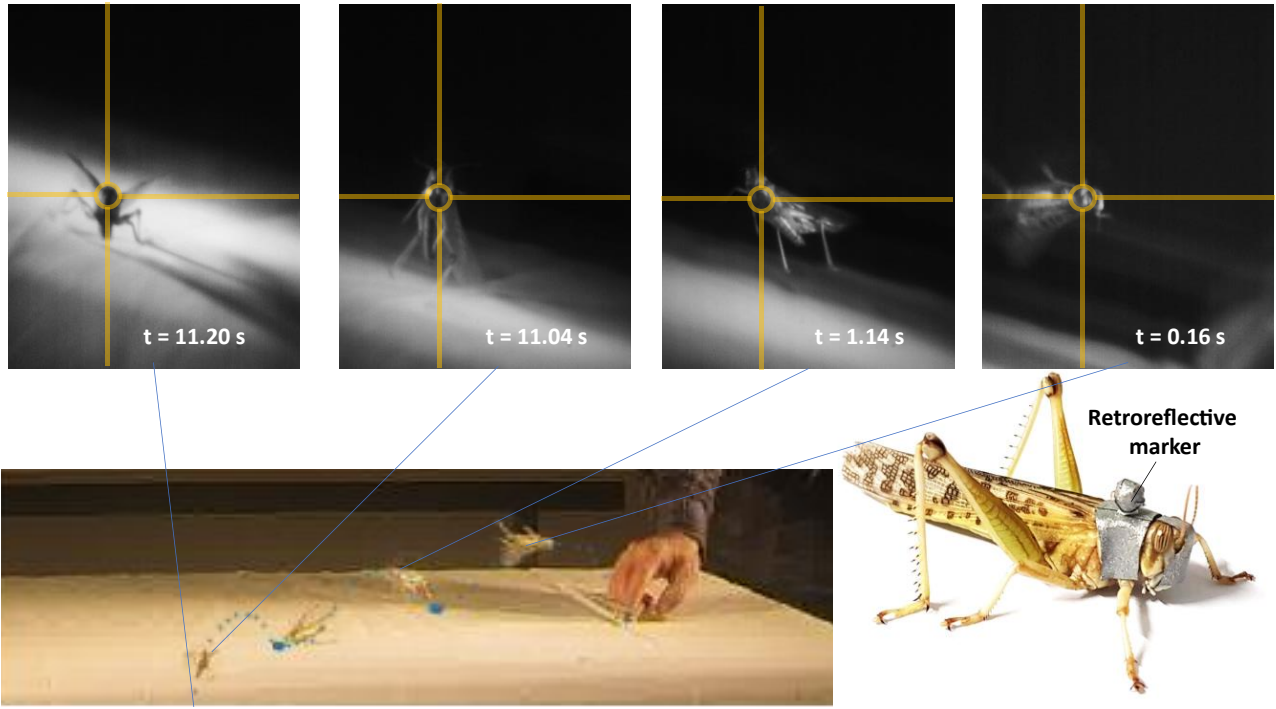

Figure 4. Multiple jumps of a locust recorded by the tracking system. The top row shows the individual frames from the high-speed telephoto camera, the bottom row shows a montage from an overview camera. Cross hairs have fixed position across frames, showing tracking stability. The retroreflective tape is wrapped around the thorax of the locust and the anterior part of the head while a 3 mm marker is mounted on the dorsal side of the thorax to make it visible from the posterior view.

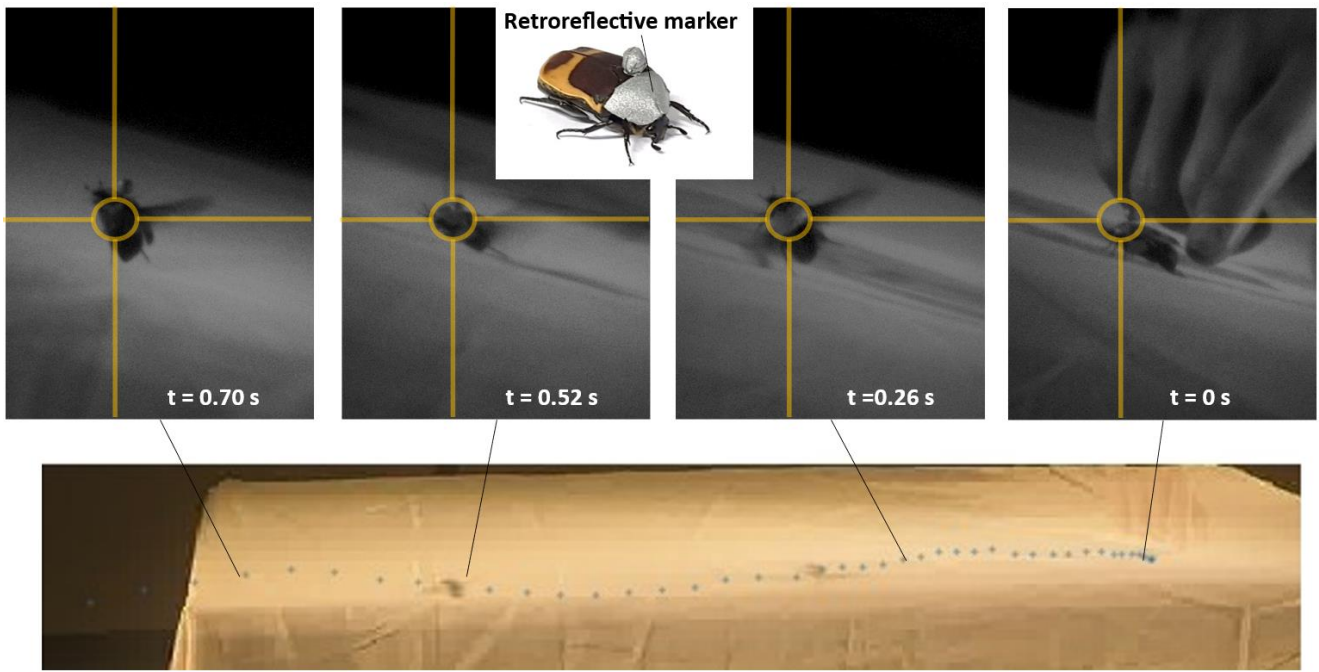

Figure 5. Taking off and flight of a beetle recorded by the tracking system. The inset shows the beetle with retroreflective tape covering its pronotum and a 3 mm marker mounted on the scutellum to increase visibility from posterior view. Individual image frames from the telephoto high speed camera are in the top row while a montage from an overview camera is shown below. Cross hairs have fixed position across frames, showing tracking stability. The beetle raised its body before taking off at t=0 s, rolled due to the wing hitting the ground (t= 0.26 s), flew straight (t= 0.52 s), and pitched up (t= 0.70 s).

## VI. Conclusion

In this paper, we tracked freely moving insects wearing reflective markers with latency on the order of one millisecond using a novel tracking system to keep an optical path aimed directly at the marker. This enabled us to make high-resolution video recordings of insects in a volume set by the angular limits of the mirrors of 40° horizontally and vertically at a distance of 3 meters. In comparison to systems based on digital image acquisition and processing, our system has low latency without sophisticated and potentially expensive digital image acquisition and processing hardware and software. While we envision many potential improvements to the present system, we suggest this system is already capable of recording videos useful for the study of insect behavior and inspiring novel innovations in bio-inspired robotics.

## Appendix

The supplemental data is at https://github.com/strawlab/msectrax and includes schematics and Gerber files of the custom circuit boards as well as the source code of the control and calibration software. The supplemental video is at https://youtu.be/zaScQgKKk3c.

## Acknowledgment

The authors thank Floris van Breugel for discussion, Tessy Balog-Albonetti for gifting the beetles. This work was supported by University of Freiburg internal funding to A.D.S. and Human Frontier Science Program Postdoctoral Cross-disciplinary Fellowship to T.T.V.-D.